\documentclass[sigconf]{acmart}

\usepackage{booktabs} 
\usepackage{array}
\usepackage{graphicx}
\usepackage{multirow}
\usepackage{slashbox}

\usepackage{color}

\setcopyright{rightsretained}

\begin{document}
\title{Lightweight Classification of IoT Malware based on Image Recognition}

\author{Jiawei Su}
\affiliation{%
  \institution{School of Information Science and Electrical Engineering, Kyushu University, Japan}
}
\email{jiawei.su@inf.kyushu-u.ac.jp}

\author{Danilo Vasconcellos Vargas}
\affiliation{%
  \institution{Faculty of Information Science and Electrical Engineering, Kyushu University, Japan}
}
\email{vargas@inf.kyushu-u.ac.jp}

\author{Sanjiva Prasad}
\affiliation{%
  \institution{Department of Computer Science and Engineering, Indian Institute of Technology Delhi, India
}
}
\email{Sanjiva.Prasad@cse.iitd.ac.in}

\author{Daniele Sgandurra}

\affiliation{%
  \institution{ISG Smart Card and IoT Security Centre, Royal Holloway University of London, UK}}
\email{daniele.sgandurra@rhul.ac.uk}

\author{Yaokai Feng}
\affiliation{%
  \institution{Faculty of Information Science and Electrical Engineering, Kyushu University, Japan}
}
\email{fengyk@ait.kyushu-u.ac.jp}

\author{Kouichi Sakurai}
\affiliation{%
  \institution{Faculty of Information Science and Electrical Engineering, Kyushu University, Japan}
}
\email{sakurai@csce.kyushu-u.ac.jp}

\begin{abstract}
The Internet of Things (IoT) is an extension of the traditional Internet, which allows a very large number of smart devices, such as home appliances, network cameras, sensors and controllers to connect to one another to share information and improve user experiences. Current IoT devices are typically micro-computers for domain-specific computations rather than traditional function-specific embedded devices. Therefore, many existing attacks, targeted at traditional computers connected to the Internet, may also be directed at IoT devices.
For example, DDoS attacks have become very common in IoT environments, as these environments currently lack basic security monitoring and protection mechanisms, as shown by the recent Mirai and Brickerbot IoT botnets.
In this paper, we propose a novel light-weight approach for detecting DDos malware in IoT environments. We firstly extract one-channel gray-scale images converted from binaries, and then utilize a light-weight convolutional neural network for classifying IoT malware families.
The experimental results show that the proposed system can achieve $94.0\%$ accuracy for the classification of goodware and DDoS malware, and $81.8\%$ accuracy for the classification of goodware and two main malware families. 
\end{abstract}

%
%

\ccsdesc[500]{Network security~ IOT network security}
\ccsdesc[300]{Machine learning~Convolutional neural network}

\keywords{Internet of things, Malware image, Convolutional neural network, Light-weight detection
}

\maketitle

\section{Introduction}
Nowadays the notion of the ``Internet'' has extended from the connection between personal computers to networks composed of a much larger range of devices. Traditional micro devices, such as many kinds of sensors and controllers, are typically only able to perform function-specific tasks based on pre-defined rules. By substituting these function-specific devices with CPU-controlled ones, and by enabling interconnection among them through the Internet, these ``things'' become ``smart'' and can now deal with complex tasks. In addition, by enabling Cloud services on these smart-devices, users can easily receive data reported by them and control them.

Despite these advantages, smarter devices imply more vulnerabilities, due to the complexity in hardware and software, with more chances for potential adversaries to threaten them. In addition, IoT systems are generally unsecured due to the difficulty of creating unified standards for the various types of IoT hardware and software platforms. Finally, even if smarter compared with traditional sensors, IoT devices still lack sufficient computational resources to be able to use existing PC-based security solutions. In some cases, Cloud services provide a way for developing security protection for IoT devices, e.g. for malware detection \cite{18, 19}.

In this paper, we consider a solution to protect 
local IoT devices from being abused to perform DDoS attacks by being enslaved in botnets of IoT devices, which is currently a common attack.
To accomplish this, we first classify 
IoT DDoS malware samples recently collected in the wild on two major families, namely Mirai and Linux.Gafgyt. We then propose a lightweight solution for detecting and classifying IoT DDoS malware and benign applications locally on the IoT devices by converting the program binaries to gray-scale images, and by feeding these images to a small size convolutional neural network for classification. In this way, resource-constrained IoT devices can afford the computation needed for running the proposed detection system locally. Experimental results show that the proposed system can achieve $94.0\%$ accuracy for classifying goodware and DDoS malware, and $81.8\%$ accuracy for the classification of goodware and two main malware families.

The main contributions of this research are the following ones:

\begin{itemize}
\item this is the first classification system tested on real IoT malware samples:  previous works have used regular or mobile malware samples instead, due to the difficulty in obtaining IoT malware samples \cite{2, 11, 13}.  Specifically there is currently no publicly available IoT malware dataset and the first IoT honeypot for collecting samples of IoT threats was released relatively recently \cite{1};

\item the IoT malware classification system can be deployed on real IoT devices. We show in detail the feasibility of using lightweight image classifier for recognizing IoT malware through malware images. Malware image classification has been proposed for classifying regular malware \cite{4}; however, IoT malware is functionally different. For example, many IoT malware may try to kill other malware to guarantee enough computational resource for themselves;

\item according to the experimental results, we prove that the proposed system can reliably classify goodware and IoT DDoS malware.

\item to the best of our knowledge, there is currently no reference to describe the time complexity of CNNs. However, the proposed CNN-based approach is empirically considered to be lightweight since it does not need to maintain any training data for classification, differently from other types of classifiers for malware, such as Support Vector Machine and K-nearest neighbours.
The computation of CNN for classification is rather simple, and only involves summation and activation. In addition, the proposed system is based on a two layer shallow network which is much more efficient than common deep learning models.
\end{itemize}

The paper is structured as follows. In Sect.~\ref{sec:related} we discuss related work. Sect.~\ref{sec:methodology} explains procedures for extracting IoT DDoS malware images and implementing a small size convolutional neural network for classification. In Sect.~\ref{sec:results} the detection results in two different scenarios are listed and in Sect.~\ref{sec:conclusion} the achievement of this research is summarized and future work is discussed.

\section{Related works}\label{sec:related}
Even if IoT security is an important topic, few defensive solutions exist in the literature \cite{12}. Only recently, the first honeypot specifically for collecting IoT malware has been established by Pa et al. \cite{1}. Their honeypot systems simulated 8 different CPU architectures and are built for observing attacks coming through the Telnet protocol. Initially they collected 43 distinct malware samples which are mostly DDoS attack malware. Their results show that the DDoS attack is the most common security threat in current IoT network environments. These authors kindly shared their observed data set with us which we have used in this research for evaluating our proposal.

To the best of our knowledge, while most other works focus on Android malware detection \cite{25, 26}, the``Cloudeye'' \cite{2} is in practice the only current work specific for IoT malware detection. The system is a signature matching-based malware detection solution. IoT clients are only responsible for preliminary scanning the software locally, and then sending hashed abstracts of suspicious files to Cloud servers for deep analysis, therefore guaranteeing data privacy and low-cost communications. However, in IoT environments the inherent weakness of signature matching-based detection still exists: for example, Cloudeye is not able to deal with new variants of existing samples.

Apart from signature matching, machine learning-based malware detection has been proved as effective in various scenarios \cite{3, 14, 15, 16, 22, 23}. In IoT environments, also heavy computation machine learning methods are expected to be suitable too because of the availability of Cloud services. In fact, in a possible scenario, the training can be performed on Cloud server, while resource-constrained IoT devices can receive the trained classifiers from the servers and run the algorithm locally. Note that several machine learning classifiers are heavy at training but efficient during test phase.

Classifying malware images has been proven as an effective way for recognizing common PC malware \cite{9, 27}. It is essentially a method for comparing two malware binaries. Nataraj et al. first utilize malware images for classifying regular Internet malware with k-nearest neighbors \cite{4}. However, the system requires pre-processing of filtering to extract the image texture as features for classification, which might not fit the resource-constrained IoT environments.  Similaly, the artificial neural network (ANN) malware classification proposed by Makandar \cite{29} might not be a good candidate to be run on IoT devices to handle due to
the heavy computational cost of multiple fully connected layers in ANN for classification. Yue utilized convolutional neural network for malware family classification \cite{5}. In this research, we use malware images for IoT malware classification and show it is a feasible approach.

\section{Methodology}\label{sec:methodology}
In this Section, we describe the methodology of feeding malware images as features, to a small two-layer convolutional neural network for detection.

\subsection{Lightweight IoT DDoS Malware Filter}

For the scenario of detecting IoT DDoS malware detection locally, as previously pointed out, the main difficulty of deploying malware filters lies in the fact that the computational resources available on current IoT devices is limited. A direct solution under such a condition is relying on the security protection services provided by powerful remote servers, such as in Cloud-enabled IoT environments.
Cloud servers are usually better protected against node failures, e.g. due to DDoS attacks. Another advantage of using Cloud servers is that a malware databases can be made more comprehensive and can be updated more rapidly than on IoT devices. For these reasons, we propose a two-tier detection architecture, based on a local IoT detection system and a remote, Cloud-based, classification system. In more detail,
firstly a lightweight malware classification system is responsible of recognizing suspicious programs and behaviors locally. Note that, at this stage, the main goal is to provide a score on binary suspiciousness.

\begin{figure}[t]
\begin{center}
\includegraphics[width=0.99\linewidth]{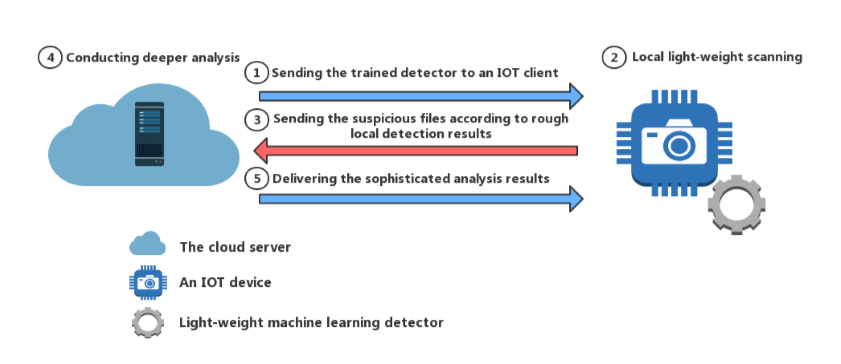}
\end{center}

   \caption{The proposed light-weight malware detection scheme. The local detector is located on the client side and captures potentially suspicious programs by relying on cloud backend for final decision \label{fig:architecture}}

\label{fig:long}
\label{fig:onecol} 
\end{figure}

In such a case, the system delivers the files or the corresponding abstracts to a remote Cloud server for deeper analysis. In addition, the cloud side can update and distribute new trained detectors to the clients periodically. In the following, we discuss the local malware filter on the client side. We assume that a set of Cloud servers are able to analyze  malware samples and retrain the classifiers using standard machine learning algorithms. The proposal system structure is shown in Fig.~\ref{fig:architecture}

\subsection{IoT DDoS Malware Families}
According to recent observations and preliminary analysis \cite{1}, even if IoT DDoS malware are functionally similar to existing DDoS malware on PC platforms, they contain specific features that are rarely observed on PCs. For example, some samples try to kill other ones of competitive families to get more system resources for themselves, due to the limited computational capability of IoT devices. In addition, IoT malware often target a wide range of devices, such as Internet cameras, DVR and so on. Finally,
IoT malware can be also compatible with different processor architectures, ensuring the maximum possible successful infections.

\subsection{Malware Image Classification}
An interesting and novel way of performing malware classification is to analyze their converted binary images. In particular, a malware binary can be reformatted as an 8-bit sequence and then be converted to a gray-scale image which has one channel and pixel values from 0 to 255 \cite{4}. The resulting image can then be fed into machine learning image classifiers for classification. 
Note that running machine learning classifiers typically needs fewer local storage than signature-matching systems, which is the most common used malware detection method. This is important for storage-constrained IoT devices. In fact, in a matching signatures system, the signature database is typically large in size as it has to contain information for each malware sample and all of its possible variants. In the case of machine learning, little information has to be kept for classification. 
For example, k-means clustering needs only the information of centroids and radii for classification once trained. Support vector machine merely keeps a small set of training data (i.e., the support vectors) in the test phase. In addition, machine learning methods overcome signature matching on detecting zero-day attacks.  Finally, converting malware binaries to the corresponding images only requires creating the input vectors to the convolutional neural network, i.e. 8-bit vectors, which is a very fast operation. 

\subsection{Neural Network for Malware Detection}
Convolutional neural networks have been proven to have better performance for image recognition than many other kinds of classifiers. A convolutional neural network has two important characteristics that make it fit the scenario of preliminary filtering malware on local IoT devices:

\begin{itemize}

\item \textbf{automatic feature extraction}: neural network can automatically extract higher level features from the input raw features. That is, the network can learn deep non-linear features that can be hardly discovered and understood by human-beings. These are sometimes actually counter-intuitive, but indeed effective. Note that many previous works have focused on extracting effective features for malware detection. However, most of them are only effective under specific scenarios, and this might lead to poor scalability.

\item \textbf{test-phase efficiency}: the training progress of a convolutional neural network requires heavy computation and, for instance, high-end graphic cards are necessary for accelerating training large networks. However, once trained, the network itself is rather lightweight and can be run with tiny computational resources, since only the trained parameters and information of network structure are kept \cite{30, 31}. In contrast, another supervised lightweight classifier, the one-class support vector machine (OCSVM), though simpler than normal Two-class SVM, still needs to keep a certain amount of training data when running the classification, while a convolutional neural network does not need to keep any. In practice, the training can be handled by the Cloud servers and only the trained network is sent to IoT nodes. On the local IoT side, the convolutional neural network is run to detect malware. 
\end{itemize}

\section{Experiment and results}\label{sec:results}
In this section we discuss the experimental setup and the results of the classification of the proposed system.

\subsection{Preparing the Dataset}
For these experiments, we have used an IoT DDoS malware dataset newly collected by IoTPOT \cite{1}, the first honeypot for collecting IoT threat samples. The malware samples are labelled using VirusTotal \cite{8} with the majority rule. The dataset originally contains 500 malware samples, where most of them are classified into four big families: Linux.Gafgyt.1, Linux.Gafgyt (other variants of Linux.Gafgyt family) 
, Mirai \cite{10} and Trojan.Linux.Fgt. The rest of the samples belong to relatively rare families such as Tsunami, Hajime, LightAidra. Then we cluster the samples into two categories: Mirai family, which contains Mirai and Trojan.Linux.Fgt\footnote{Mirai has been shown to have similar features to Trojan.Linux.Fgt \cite{17}.}, and Linux.Gafgyt family which contains Linux.Gafgyt.1 and the other variants.
Instead, the benign binary samples (goodware) are collected from Ubuntu 16.04.3 system files. The number of samples are balanced for each family by randomly removing the samples that belong to classes that are too large. After the preprocessing phase, we analyzed 365 samples where each class has the same number of samples. Among them, we utilize 45 sample (each class has 15 samples) for testing, and the rest for training. According to the discussion above, the system proposed is only responsible for preliminary detection. That is, the goal is to identify whether a sample is benign or belongs to one of the big malware families: Mirai and Linux.Gafgyt, but there is no need to understand exactly which kind of variant it is.

\subsection{Obtaining the Malware Images}
\begin{figure}[t]
\begin{center}
\includegraphics[width=0.8\linewidth]{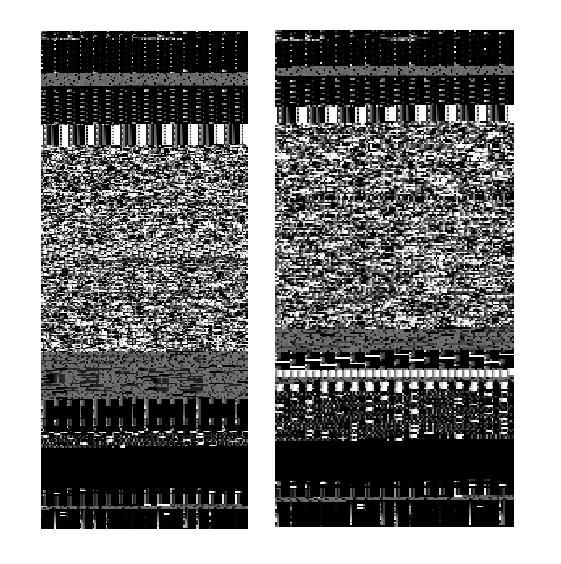}
\end{center}
\vspace{3mm}
\begin{center}
\includegraphics[width=0.6\linewidth]{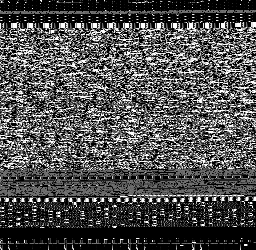}
\end{center}

   \caption{Images of Goodware \label{fig:goodware1}}
\label{fig:long}
\label{fig:onecol} 
\end{figure}

\begin{figure}[t]
\begin{center}
\includegraphics[width=0.5\linewidth]{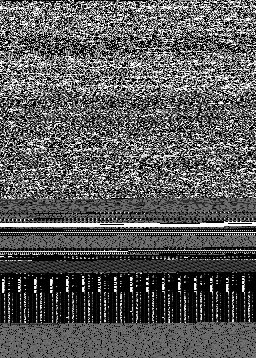}
\end{center}
\vspace{3mm}
\begin{center}
\includegraphics[width=0.5\linewidth]{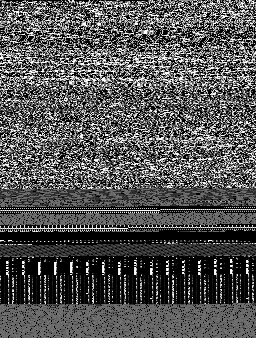}
\end{center}
\vspace{3mm}
\begin{center}
\includegraphics[width=0.5\linewidth]{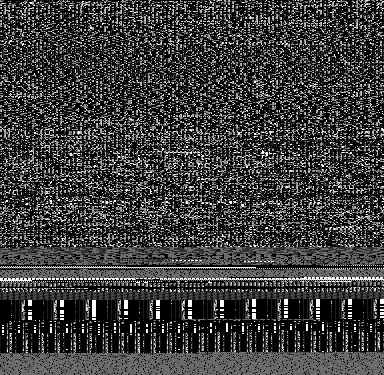}
\end{center}
   \caption{Malware Image Examples of the Linux.Gafgyt Family \label{fig:goodware2}}
\label{fig:long}
\label{fig:onecol}
\end{figure}

\begin{figure}[t]
\begin{center}
\includegraphics[width=0.8\linewidth]{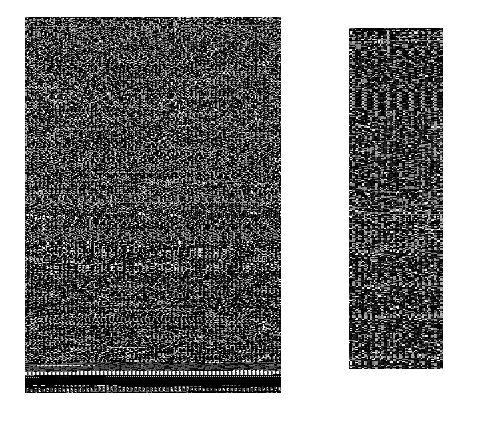}
\end{center}
\vspace{3mm}
\begin{center}
\includegraphics[width=0.5\linewidth]{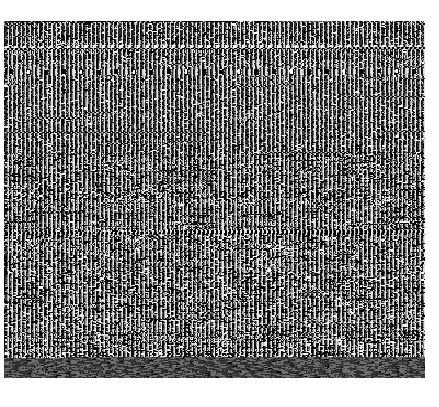}
\end{center}
\vspace{3mm}
\begin{center}
\includegraphics[width=0.5\linewidth]{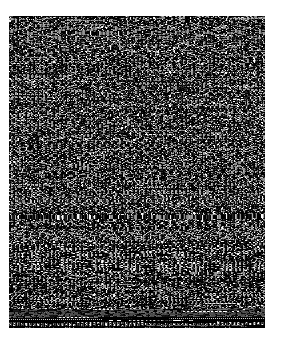}
\end{center}
   \caption{Malware Image Examples of the Mirai Family \label{fig:goodware3}}
\label{fig:long}
\label{fig:onecol}
\end{figure}

We then convert each sample of the dataset into the corresponding malware gray-scale image by following the same procedures implemented in \cite{4}. In particular, a malware binary can be reformatted as an 8-bit string sequence, whose decimal encoding represents the value of a one-channel pixel (in the range $[0, 255]$).
Therefore the entire sequence represents a gray-scale image.  We rescale the images to the size of 64X64 to be used as input to a convolutional neural network. Some examples of malware and benign-ware images are shown by Fig.~\ref{fig:goodware1}, \ref{fig:goodware2} and \ref{fig:goodware3}. In these images, the structural difference between malware and goodware images can be easily identified. For example, it can be seen that malware images always are more dense. In particular, the majority of the Mirai malware images have a dense central code payload. On the other hand, the images of goodware tend to have larger header parts than malwares.

\vspace{3mm}

\subsection{Convolutional Neural Network Configuration}
To have a lightweight detection system, we have implemented a small, two layer shallow convolutional neural network, compared with common image recognition models, such as ImageNet \cite{20} and VGG \cite{7}. The network structure is shown in Table.~\ref{fig:conv_network}. The network is trained with 5000 iterations with a training batch size of 32 and learning rate 0.0001. 

\begin{table}
\begin{center}
\begin{tabular}{|c|c|}
\hline
convolution layer(kernel=3, stride=1, depth=32) \\
max pooling layer(kernel=2, stride=2) \\
convolution layer(kernel=3, stride = 1, depth=72) \\
max pooling layer(kernel=2, stride=2) \\
fully connected layer(size=256) \\
softmax classifier \\
\hline
\end{tabular}
\end{center}
\caption{Structure of the Implemented Convolutional Neural Network \label{fig:conv_network}}
\end{table}

\subsection{Results}

The classification results are shown in Tables ~\ref{table:3-class-confusion} and ~\ref{table:2-class-confusion} for the cases of two (benign and malicious) and three-class (benign and two malware families: Mirai and gafgyt) classification. The experiments were conducted five times which each time with a completely different training/test data combination (i.e., there are no shared test samples between any two of five test data sets).

\begin{table}[!hbt]
\begin{center}
\begin{tabular}{|c|c|c|c|}
	\hline \backslashbox{True}{Predict} & Benign & Gafgyt & Mirai\\
	\hline Benign&$94.67\%$&$2.67\%$&$2.67\%$\\
	\hline Gafgyt&$6.67\%$&$72.00\%$&$21.33\%$\\
	\hline Mirai&$0\%$&$21.33\%$&$78.67\%$\\
	\hline
\end{tabular}
\end{center}
\caption{Confusion Matrix for 3-class Classification \label{table:3-class-confusion}}
\end{table}

\begin{table}[!hbt]
\begin{center}
\begin{tabular}{|c|c|c|}
	\hline \backslashbox{True}{Predict} & Benign & Malicious\\
	\hline Benign&$94.67\%$&$5.33\%$\\
	\hline Malicious&$6.67\%$&$93.33\%$\\

	\hline
\end{tabular}
\end{center}
\caption{Confusion Matrix for 2-class Classification \label{table:2-class-confusion}}
\end{table}

\begin{table}[!hbt]
\begin{center}
\begin{tabular}{|c|c|c|c|}
	\hline Class & Goodware & Gafgyt & Mirai\\
	\hline Time consumption in second&0.0241&0.0011&0.0003\\

	\hline
\end{tabular}
\end{center}
\caption{Practical results of time complexity for classifying one image of goodware and two malware families.  \label{tab:time}}
\end{table}

\begin{table*}[!hbt]
\centering
\begin{center}
\begin{tabular}{|c|c|c|c|}
	\hline \backslashbox{Metrics}{Systems} & Our method & ANN with random projection \cite{5}&Weighted loss \cite{3}\\
	\hline Accuracy &$94.0\%$&$99.5\%$&$96.9\%$\\
	\hline Classifier &CNN&ANN&CNN (VGG-s)\\
	\hline Num of layers &2&2&5\\
	\hline Num of nodes &104&1536&1888\\
	\hline Fully connect layer &256&2048&4096X2\\
	\hline Preprocess &Re-organizing binary&N-gram binary, Random projection&Re-organizing binary\\
	\hline Input dimension &64X64 scalar matrix&179 thousand binaries&Unknown\\
	\hline
\end{tabular}
\end{center}
\caption{Comparing proposed system with two previous related works. In particular, the number of hidden layers, number of neurons and the number of nodes in fully connect layers, are shown by ``Num of Layers'',``Num of nodes'',``Fully connect layer'' respectively. It can be seen that the proposed system is more lightweight than references due to the smaller size of network model and lower dimensions of input, as well as simpler preprocessing. \label{tab:comparison}}
\end{table*}

According to the results of two-class classification, we find the proposed system can  predict the existence of maliciousness with about $94.0\%$ accuracy on the average. The accuracy of three-class classification is relatively lower. Specifically, there are $6.67\%$ malicious samples that are mis-classified as benign, all of which belong to Gafgyt family, while there is no misclassification of Mirai family to benign. This indicate the Gafgyt has more similar binaries to goodware. On the other side, the rate of misclassification between Mirai and Gafgyt is exactly the same.
Generally, the difference between benign and malicious samples is more recognizable than the difference between two malware families. Comparing with misclassification between benign and malicious samples (i.e., two-class classification), the system is more likely to misclassify the samples of two malware families in the case of three-class classification. This indicates the similarity between these two families. Specifically, samples of two families might be obfuscated in similar ways, or/and share a part of the malicious functions. In fact, the basic botnet functions of different DDoS malware are similar, and mainly include receiving instructions from the control server and spreading the infection. Also consider that Mirai source code has been available online since its beginning, and several new families of IoT botnets include some portions of Mirai code. In addition, IoT malware has to be lightweight and their functions have to be relatively direct and simple.


Our accuracy results compete with similar previous works \cite{3, 5}. In specific, Yue \cite{5} also utilized convolutional neural networks and malware images for classifying several PC malware families. However the results are carried out by using much bigger and complex network structures, namely very deep networks (VGG) which contain more than 10 layers while ours only has two layers. Similarly, a very complex preprocess procedure is needed in \cite{5} which involves initial feature selection and random projection while our proposal directly uses raw features for classification. According to the accuracy results, the proposed system can be utilized as a regular malware detector, or a first-stage malware classifier. That is, it can perform a precise classification to identify benign and maliciousness .The exact classification of the malware family can then be performed on a Cloud backend.
A comparison of corresponding experimental accuracy and settings is shown by Table~\ref{tab:comparison}.
The practical time cost of classifying images is depicted in Table~\ref{tab:time}. In detail, the experiment is conducted with python and tensorflow on a system running Ubuntu 14.04, with a i7-7700 processor, GTX1080Ti graphic card and 16G memory. The code of this research can be found in: https://github.com/Carina02/IotMalwareImage.

\section{Conclusion and Future work}\label{sec:conclusion}
In this paper we have proposed a lightweight malware image classification scheme for locally detecting IoT DDoS malware, and shown its feasibility. The malware filter proposed in this paper is based on convolutional neural networks and can be tuned to be more efficient by using various techniques of reducing network size. For example, removing the neurons and links that are not critical in the network can reduce the number of parameters needed for classification \cite{21}. Such further optimization can make the proposed system implementable on IoT devices with even less computation resources. In addition, new malware image extraction methods can be proposed to obtain more representative features of malware for classification.

For improving the detection rate of IoT malware, more complicated cases in practice can be considered such as malware obfuscation. To the best knowledge, there is currently no systematical evaluation on IoT malware obfuscation and several critical questions are yet to be answered, such as whether IoT malware is obfuscated in a similar way to traditional malware, and how limited resources influence obfuscation methods.





\begin{acks}
This research was partially supported by Collaboration Hubs for International Program (CHIRP) of SICORP, Japan Science and Technology Agency (JST), and Project of security in the IoT space funding by Department of Science and Technology (DST), India.

\end{acks}


\begin{thebibliography}{}
\bibitem{1} Pa, Y.M.P., Suzuki, S., Yoshioka, K., Matsumoto, T., Kasama, T. and Rossow, C., 2015. IoTPOT: analysing the rise of IoT compromises. EMU, 9, p.1.
\bibitem{2} Sun, H., Wang, X., Buyya, R. and Su, J., 2017. CloudEyes: Cloud based malware detection with reversible sketch for resource constrained internet of things (IoT) devices. Software: Practice and Experience, 47(3), pp.421-441.
\bibitem{3} Dahl, G.E., Stokes, J.W., Deng, L. and Yu, D., 2013, May. Large-scale malware classification using random projections and neural networks. In Acoustics, Speech and Signal Processing (ICASSP), 2013 IEEE International Conference on (pp. 3422-3426). IEEE.
\bibitem{4} Nataraj, L., Karthikeyan, S., Jacob, G. and Manjunath, B.S., 2011, July. Malware images: visualization and automatic classification. In Proceedings of the 8th international symposium on visualization for cyber security (p. 4). ACM.
\bibitem{5} Yue, S., 2017. Imbalanced Malware Images Classification: a CNN based Approach. arXiv preprint arXiv:1708.08042.
\bibitem{6} LeCun, Y., 2015. LeNet-5, convolutional neural networks. URL: http://yann. lecun. com/exdb/lenet.
\bibitem{7} Simonyan, K. and Zisserman, A., 2014. Very deep convolutional networks for large-scale image recognition. arXiv preprint arXiv:1409.1556.
\bibitem{8} Https://www.virustotal.com/
\bibitem{9} Wagner, M., Fischer, F., Luh, R., Haberson, A., Rind, A., Keim, D.A., Aigner, W., Borgo, R., Ganovelli, F. and Viola, I., 2015. A survey of visualization systems for malware analysis. In EG Conference on Visualization (EuroVis)-STARs (pp. 105-125).
\bibitem{10} Kolias, C., Kambourakis, G., Stavrou, A. and Voas, J., 2017. DDoS in the ioT: Mirai and other botnets. Computer, 50(7), pp.80-84.
\bibitem{11} Alam, M.S. and Vuong, S.T., 2013, August. Random forest classification for detecting android malware. In Green Computing and Communications (GreenCom), 2013 IEEE and Internet of Things (iThings/CPSCom), IEEE International Conference on and IEEE Cyber, Physical and Social Computing (pp. 663-669). IEEE.
\bibitem{12} Zhang, Z.K., Cho, M.C.Y., Wang, C.W., Hsu, C.W., Chen, C.K. and Shieh, S., 2014, November. IoT security: ongoing challenges and research opportunities. In Service-Oriented Computing and Applications (SOCA), 2014 IEEE 7th International Conference on (pp. 230-234). IEEE.
\bibitem{13} Ham, H.S., Kim, H.H., Kim, M.S. and Choi, M.J., 2014. Linear SVM-based android malware detection for reliable IoT services. Journal of Applied Mathematics, 2014.
\bibitem{14} Firdausi, I., Erwin, A. and Nugroho, A.S., 2010, December. Analysis of machine learning techniques used in behavior-based malware detection. In Advances in Computing, Control and Telecommunication Technologies (ACT), 2010 Second International Conference on (pp. 201-203). IEEE.
\bibitem{15} Ahmed, F., Hameed, H., Shafiq, M.Z. and Farooq, M., 2009, November. Using spatio-temporal information in API calls with machine learning algorithms for malware detection. In Proceedings of the 2nd ACM workshop on Security and artificial intelligence (pp. 55-62). ACM.
\bibitem{16} Shamili, A.S., Bauckhage, C. and Alpcan, T., 2010, August. Malware detection on mobile devices using distributed machine learning. In Pattern Recognition (ICPR), 2010 20th International Conference on (pp. 4348-4351). IEEE.
\bibitem{17} Hallman, R., Bryan, J., Palavicini, G., Divita, J. and Romero-Mariona, J., 2017. IoDDoS The Internet of Distributed Denial of Sevice Attacks.
\bibitem{18} Burguera, I., Zurutuza, U. and Nadjm-Tehrani, S., 2011, October. Crowdroid: behavior-based malware detection system for android. In Proceedings of the 1st ACM workshop on Security and privacy in smartphones and mobile devices (pp. 15-26). ACM.
\bibitem{19} Masud, M.M., Al-Khateeb, T.M., Hamlen, K.W., Gao, J., Khan, L., Han, J. and Thuraisingham, B., 2011. Cloud-based malware detection for evolving data streams. ACM transactions on management information systems (TMIS), 2(3), p.16.
\bibitem{20} Krizhevsky, A., Sutskever, I. and Hinton, G.E., 2012. Imagenet classification with deep convolutional neural networks. In Advances in neural information processing systems (pp. 1097-1105).
\bibitem{21} Pan, W., Dong, H. and Guo, Y., 2016. DropNeuron: Simplifying the Structure of Deep Neural Networks. arXiv preprint arXiv:1606.07326.
\bibitem{22} Shabtai, A., Moskovitch, R., Elovici, Y. and Glezer, C., 2009. Detection of malicious code by applying machine learning classifiers on static features: A state-of-the-art survey. information security technical report, 14(1), pp.16-29.
\bibitem{23} Siddiqui, M., Wang, M.C. and Lee, J., 2008, March. A survey of data mining techniques for malware detection using file features. In Proceedings of the 46th annual southeast regional conference on xx (pp. 509-510). ACM.
\bibitem{25} Schmidt, A.D., Bye, R., Schmidt, H.G., Clausen, J., Kiraz, O., Yuksel, K.A., Camtepe, S.A. and Albayrak, S., 2009, June. Static analysis of executables for collaborative malware detection on android. In Communications, 2009. ICC'09. IEEE International Conference on (pp. 1-5). IEEE.
\bibitem{26} Felt, A.P., Finifter, M., Chin, E., Hanna, S. and Wagner, D., 2011, October. A survey of mobile malware in the wild. In Proceedings of the 1st ACM workshop on Security and privacy in smartphones and mobile devices (pp. 3-14). ACM.
\bibitem{27} Makandar, A. and Patrot, A., 2018. Trojan Malware Image Pattern Classification. In Proceedings of International Conference on Cognition and Recognition (pp. 253-262). Springer, Singapore.
\bibitem{28} Makandar, A. and Patrot, A., 2015. Overview of malware analysis and detection. In IJCA proceedings on national conference on knowledge, innovation in technology and engineering, NCKITE (Vol. 1, pp. 35-40).
\bibitem{29} Makandar, A. and Patrot, A., 2015, December. Malware analysis and classification using Artificial Neural Network. In Trends in Automation, Communications and Computing Technology (I-TACT-15), 2015 International Conference on (Vol. 1, pp. 1-6). IEEE.
\bibitem{30} Wang, S.C., 2003. Artificial neural network. In Interdisciplinary computing in java programming (pp. 81-100). Springer, Boston, MA.
\bibitem{31} Krizhevsky, A., Sutskever, I. and Hinton, G.E., 2012. Imagenet classification with deep convolutional neural networks. In Advances in neural information processing systems (pp. 1097-1105).

\end{thebibliography}
{}
\end{document}